# Taylor-SWFT: fast discrete Statistical Wave Field Theory using Taylor expansion for late reverberation

WORK UNDER REVIEW

Marius RODRIGUES
*LTCI, Télécom Paris*
marius.rodrigues@telecom-paris.fr

Louis LALAY
*LTCI, Télécom Paris*
louis.lalay@telecom-paris.fr

Roland BADEAU
*LTCI, Télécom Paris*
roland.badeau@telecom-paris.fr

Gaël RICHARD
*LTCI, Télécom Paris*
gael.richard@telecom-paris.fr

Mathieu FONTAINE
*LTCI, Télécom Paris*
mathieu.fontaine@telecom-paris.fr

*Abstract*—Dynamic room acoustic simulation aims to render the acoustic effects of an environment in real time while accounting for potentially moving sources and receivers. In this context, the efficient synthesis of the long-term room response, also known as late reverberation, remains challenging because of the intricate relationship between room geometry and acoustic behavior. This paper introduces Taylor-SWFT, an efficient implementation of key results from Statistical Wave Field Theory (SWFT) for the geometry-aware dynamic synthesis of late reverberation. The method is evaluated on the Benchmark for Room Acoustical Simulation (BRAS) and achieves competitive performance compared with classical approaches, while substantially reducing computational cost.

## I. INTRODUCTION

Acoustic field simulation refers to the realistic numerical rendering of sound wave propagation within a given environment. A particular subset of these methods addresses *room acoustic simulation*, also known as *reverberation synthesis*. A large variety of sound rendering techniques have been designed for this purpose [1]. In contrast, dynamic acoustic simulation concerns the real-time rendering of acoustic responses in environments involving potentially moving sound sources and receivers. This paper's focus is placed on *dynamic room acoustic simulation* with mono-channel source and receiver. Typical applications are found in the entertainment industry, particularly in video games and virtual reality [2], but also extend to domains such as hearing aids [3], teleconferencing systems [4] and a range of downstream tasks including data augmentation for machine learning, acoustic transfer, and dynamic speech enhancement. Within this context, established techniques such as the Image Source Method (ISM), Ray Tracing (RT) [5] and Acoustic Radiance Transfer (ART) [6] are generally unsuitable due to their potentially high computational cost. A widely adopted strategy for addressing this problem consists in decomposing the reverberation effect into two parts, namely the *early echoes* and the *late reverberation*, and subsequently applying dedicated models to synthesize each component [7]. Recent works also explored data-driven approaches in which neural networks are trained to approximate the sound field from a set of acoustic measurements [8], [9], [10].

Recently introduced, the Statistical Wave Field Theory (SWFT) [11] provides a physics-based, statistical, spatiotemporal description of late reverberation, derived from the asymptotic analysis of the wave equation in the regime of long times and high frequencies. Whereas this theory provides an accurate acoustic field representation, the formulation as provided in [11] is computationally demanding. In the present work, we introduce Taylor-SWFT, a fast dynamic reverberator that combines a novel SWFT-based model for late reverberation with a low-order ISM for early echoes. The contribution of this paper is twofold:

- First, we propose a new geometry-aware, physics-based late reverberation synthesizer capable of dynamically adapting to variations in the receiver position.
- Second, we develop an efficient implementation of the SWFT based on a Taylor expansion. The method exhibits reduced latency and low initialization time, which allows its application in real-time scenarios.

The paper is structured as follows. Section II presents an overview of the SWFT and its major theoretical results. In Section III, we derive a discrete formulation of these results and deduce a fast Room Impulse Response (RIR) generation method. Finally, Section IV provides a comprehensive evaluation on the BRAS dataset (Benchmark for Room Acoustical Simulation) [12], along with a discussion of the performance.

**Notations**
- $h(x,t)$: RIR as a function of space $x \in \mathbb{R}^3$ and time $t > 0$;
- $\mathcal{F}_t\{h(x,t)\}$: Fourier transform of $h$ with respect to parameter $t$;

- $\gamma_h(x_1, x_2, t_1, t_2) = \mathbb{E}[h(x_1, t_1) h^*(x_2, t_2)]$: spatio-temporal autocovariance of $h$;
- $W_h(x_1, x_2, t, \cdot) = \mathcal{F}_\tau\{\gamma_h(x_1, x_2, t + \frac{\tau}{2}, t - \frac{\tau}{2})\}$: Wigner-Ville distribution of $h$;
- $V$, $\partial V$: respectively a bounded domain in $\mathbb{R}^3$ and its boundary surface;
- $|V|$, $|\partial V|$: respectively the volume and area of $V$ and $\partial V$;
- $a(s, f)$: boundary absorption at position $s \in \partial V$ and frequency $f$;
- $c = 343\ m.s^{-1}$: speed of sound.

## II. Statistical Wave Field Theory

From a physical point of view, reverberation is the diffuse mixing of a sound wave with its echoes against the room's walls. From the perspective of signal processing, this can be considered as a linear filtering operation, characterized by the RIR. Recently introduced, the SWFT proposes a probabilistic approach for solving the wave equation [11] under Neumann's and Robin's boundary conditions. Its results generalize those from [13], allowing to derive the spatio-temporal first and second moments of any RIR, in the asymptotic regime of high frequencies and long durations. With formal words, under reasonable acoustic conditions on the room's shape $V$ (mixing room) and materials (local reaction), late reverberation in $V$ is statistically described as follows:

- $h(x, t)$, the RIR at spatial position $x$ and time $t > 0$ is Gaussian and centered;
- its spatio-temporal Wigner-Ville distribution can be factorized in the form:

$$W_h(x_1, x_2, t, f) = B(x_1, x_2, f) e^{-\alpha(f)t}, \quad (1)$$

where $\alpha$ is an even and strictly positive function, and $B$ is strictly positive and even w.r.t. $f$ when $x_1 = x_2$. This distribution can be interpreted as the covariance of the wave field between two spatial positions $x_1$ and $x_2$, at fixed frequency $f$ and time $t$. A nice feature of the SWFT is that it provides explicit formulas for $\alpha$ and $B$ (Eqs. (123) and (124) in [11]). Specifically, under a first-order asymptotic expansion and considering $x := x_1 = x_2$, the following expressions hold:

$$\alpha(f) = \frac{c}{4|V|} \int_{\partial V} \ln(1 - a(s, f)) dS(s), \quad (2)$$

and

$$B_x(f) = \bar{B}(f) \int_V \operatorname{sinhc}\left(\frac{2\alpha(f)}{c} \|x - v\|_2\right) dv, \quad (3)$$

with

$$\bar{B}(f) := \frac{cf^2}{\pi\ |V|^2 \left[(2f)^2 + \left(\frac{\alpha(f)}{2\pi}\right)^2\right]}. \quad (4)$$

Here, $a(s, f)$ represents the walls' absorption (its expression is different from the usual one, which is often referred to as the 'Paris formula', and it is provided in Eqs. (125) and (126) in [11]), $c$ is the speed of sound, $B_x(f)$ denotes $B(x, x, f)$ and $\operatorname{sinhc}(x) := \frac{\sinh(x)}{x}$. Remarkably, (2) is consistent with Eyring's formula, and $\alpha$ is not dependent on the listener's position. The accuracy of Eq. (2) has been confirmed by numerical experiments in [14], whereas Eq. (3) holds on average when the source position is randomly distributed in the room.

This paper aims to provide a feasible numerical implementation of these equations and to allow fast sampling of impulse responses $h$ whose statistics over time are consistent with (1). Our proposed model, Taylor-SWFT, combines SWFT-accurate modeling of late reverberation with a low-order ISM for early echoes, enabling real-time single-channel reverberation with a moving source and receiver. Finally, it is important to note that spatial correlations are not considered in our formulation, so a sampled impulse response cannot be interpreted as a valid *spatial* RIR. Nonetheless, the SWFT does account for spatial correlations; including them is left for future work.

## III. Towards discrete implementation

The goal of this section is to provide implementable discrete versions of (1), (2) and (3).

### A. Covariance matrix

Let $h_x$ be a non-stationary continuous Gaussian random process satisfying (1) (with $x_1 = x_2 = x$). Then, its autocovariance writes as the inverse Fourier transform of its Wigner-Ville distribution ((1)):

$$\gamma_{h_x}(t_1, t_2) = \int_{\mathbb{R}} B_x(f) e^{-\alpha(f)\frac{t_1+t_2}{2} + 2i\pi f(t_1 - t_2)} df. \quad (5)$$

The discrete version of this formula is expressed in terms of the inverse Discrete Fourier Transform (DFT), and evaluated at $N$ discrete time steps $t = \frac{n}{F_s}$, with $F_s$ the sampling frequency. This provides a covariance matrix $\Sigma_x$ whose term $(n, m)$ is:

$$\Sigma^x_{nm} = \frac{1}{N} \sum_{k=-\lfloor \frac{N}{2} \rfloor}^{\lfloor \frac{N}{2} \rfloor - 1} B_x(f_k) e^{-\alpha(f_k)\frac{n+m}{2F_s} + \frac{2i\pi k(n-m)}{N}}, \quad (6)$$

with $f_k = \frac{kF_s}{N}$. Then, using the isometry property of the DFT, it can be shown that (6) can be expressed in terms of a scalar product:

$$\Sigma^x_{nm} = \frac{1}{F_s^2} \langle R^x_n, R^x_m \rangle, \quad (7)$$

where $R^x_n$ is the column vector whose $s$-th term is $R^x_{s,n} = g_x \star p^{\star n}[s - n]$ with $g_x = \operatorname{DFT}_k^{-1}\{\sqrt{B_x(f_k)}\}$, and $p = \operatorname{DFT}_k^{-1}\{e^{-\frac{\alpha(f_k)}{2F_s}}\}$. Here, $\star$ denotes the convolution product and $p^{\star n} := \underbrace{p \star \cdots \star p}_{n \text{ times}}$. Then, (7) can be written in matrix form:

$$\Sigma_x = \frac{1}{F_s^2} R_x^T R_x. \quad (8)$$

Denoting $G_x$ the Toeplitz matrix generated from $g_x$ and $P$ the matrix whose $n$-th column is $p^{\star n}[\cdot -n]$, $R_x$ rewrites as

$$R_x = G_x P. \quad (9)$$

In the end, $h_x$ can be simulated in the discrete time domain with

$$\hat{h}_x = \frac{1}{F_s} R_x^T \varepsilon = \frac{1}{F_s} P^T G_x^T \varepsilon, \quad (10)$$

where $\varepsilon$ is centered Gaussian white noise of variance 1. From a physical point of view, $G_x^T$ models the stationary effect of the room at position $x$, and $P^T$ is a coloring operator generating a frequency-dependent exponential energy decay. Remarkably, (10) is just one transpose operation away from the reverberation model that was previously proposed in [15]:

$$\tilde{h}_x = \frac{1}{F_s} R_x \varepsilon = \frac{1}{F_s} G_x P \varepsilon. \quad (11)$$

In practice, we observe that both models provide very close results (Fig. 1). This leads us to consider the following class of parameters:

$$\mathcal{R} = \{(p, g_x) \mid \exists O \in \Omega(\mathbb{R}^\mathbb{N}); P^T G_x^T = O G_x P\}, \quad (12)$$

where $\Omega(\mathbb{R}^\mathbb{N})$ is the set of $N \times N$ orthogonal matrices. In that class, both Gaussian vectors $G_x P \varepsilon$ and $P^T G_x^T \varepsilon$ follow the same probability distribution $\mathcal{N}(0, \Sigma_x)$. From a computational point of view, (11) is more efficient than (10), as discussed in the next section.

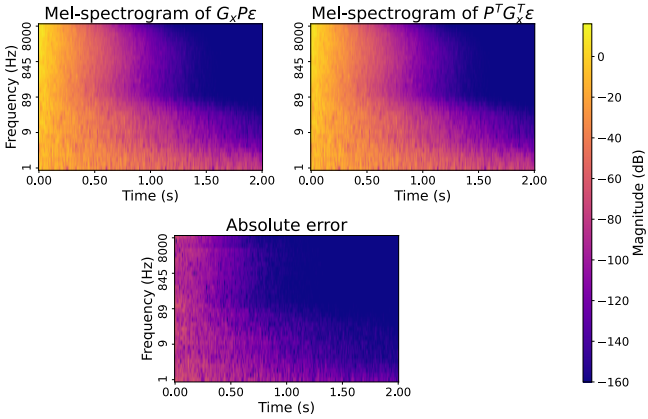

Fig. 1. Mel-Spectrogram comparison of two RIRs generated with either (10) or (11)

### B. Real-time inference

At inference time, the proposed algorithm applies late reverberation to a given source signal $s$ for a microphone placed at position $x$ in the following fashion. Let $P$, $G_x$ and $g_x$ be defined as in Section III.A, and $\bar{h} := P\varepsilon$. Using (11), the resulting reverberated signal is:

$$y_x = \hat{h}_x \star s = \frac{1}{F_s}(G_x P \varepsilon) \star s = \frac{1}{F_s} \bar{h} \star g_x \star s. \quad (13)$$

Remarkably, under the assumption that $B_x$ is smooth, $g_x$ can be designed as a short-term finite impulse response filter, containing only a few non-zero coefficients. Consequently, considering a moving position $x$, updates of $g_x \star s$ are fast, with a buffer size only depending on the evaluation speed of $x \mapsto B_x$. Moreover, $\bar{h}$ is independent from $x$; it only needs to be computed once at the room's instantiation time. Note that this optimization wouldn't have been possible if we used (10) instead of (11) to generate the RIR.

### C. Taylor expansion based fast coloring

While the colored energy decay $\bar{h} = P\varepsilon$ is only generated once, an efficient implementation of this operation is important to avoid long loading times. Previous works in [16] provided a fast algorithm in that aim, reducing the computation time to a few milliseconds. In a nutshell, the method consists in the following. Denoting $\bar{H}$, $E$ and $\mathcal{P}$ the z-transforms of $\bar{h}$, $\varepsilon$ and $p$, it can be shown that

$$\begin{aligned} \bar{H}(e^{2i\pi f}) &= E\left(\frac{e^{2i\pi f}}{\mathcal{P}(e^{2i\pi f})}\right) \\ &= \sum_{n=0}^{N} \varepsilon_n \left(\frac{\mathcal{P}(e^{2i\pi f})}{e^{2i\pi f}}\right)^n. \end{aligned} \quad (14)$$

Equivalently, the Fourier transform of $\bar{h}$ is obtained by evaluating the polynomial $A(z) := \sum_{n=0}^{N} \varepsilon_n z^n$ at $z_f = \frac{\mathcal{P}(e^{2i\pi f})}{e^{2i\pi f}} = e^{-\frac{\alpha(f)}{2F_s} - 2i\pi f}$. Performing this evaluation for all frequency bins $f_1, ..., f_N$ has a complexity of $O(N^2)$, which is not efficient given that the desired output length $N$ is usually very large. To address this issue, the method proposed in [16] considers the Taylor expansion of $A$ around $\bar{z}_f = e^{-\frac{\bar{\alpha}}{2F_s} - 2i\pi f}$ where $\bar{\alpha}$ is a user-defined parameter:

$$A(z_f) = \sum_{m=1}^{M} \frac{A^{(m)}(\bar{z}_f)}{m!} (z_f - \bar{z}_f)^m. \quad (15)$$

Here, as long as $\bar{\alpha}$ is close to $\alpha(f)$ for all $f$, $M$ can be chosen much smaller than $N$. Moreover, further computations show that all coefficients $A^{(m)}(\bar{z}_f)$ can be obtained through a Fast Fourier Transform (FFT). In the end, using the Taylor expansion in (15) leads to a complexity of $O(MN \log(N))$. Finally, a good choice for $\bar{\alpha}$ is

$$e^{-\frac{\bar{\alpha}}{2F_s}} = \frac{\max_f e^{-\frac{\alpha(f)}{2F_s}} + \min_f e^{-\frac{\alpha(f)}{2F_s}}}{2}. \quad (16)$$

Indeed, this choice makes a trade-off between efficiency (i.e. lower acceptable values for the Taylor expansion's order $M$) and stability ($z_f$ stays in the convergence disk of the extended power series $\sum_{m \geq 0} \frac{A^{(m)}(\bar{z}_f)}{m!}(z - \bar{z}_f)^m$).

### D. Reverberation parameters

(2) and (3) provide a way to compute reverberation parameters directly from the room's shape and the material. The implementation of these two formulas is rather direct and is done by approximating the integrals with Riemann sums. In practice for (2), we consider a triangulation $\mathcal{T} = \{T_1, ..., T_K\}$ of the room's surface, such that $a(s, f)$ is almost constant on each triangle $T_i$. We denote $\alpha_i(f) := \alpha(s, f)$ when $s \in T_i$. Then the integral over $\partial V$ in (2) is replaced by a sum over wall elements $T_i$:

$$\hat{\alpha}(f) = \frac{c}{4|V|} \sum_{i=1}^{K} |T_i| \ln(1 - a_i(f)). \quad (17)$$

Similarly for (3), the room's volume is divided into $L$ cubic voxels $\mathcal{V} = \{V_1, ..., V_L\}$ with centers $v_1, ..., v_L \in \mathbb{R}^3$. Then, $B_x$ is approximated by:

$$\hat{B}_x(f) = \bar{B}(f) \sum_{i=1}^{L} |V_i| \operatorname{sinhc}\left(\frac{2\alpha(f)}{c} \|x - v_i\|_2\right), \quad (18)$$

where $\bar{B}$ was defined in (4).

(17) and (18) might look computationally heavy, especially when dealing with complex rooms that need fine-grained triangulation. To enable real-time implementation, we note that a fair amount of the computations are only needed once. For instance, (17) only depends on the room's shape and the materials, which are considered constant over time. On the other hand, (18) depends on the receiver's position $x$, which may vary over time. Fortunately, $x \mapsto \hat{B}_x(f)$ is a smooth and strictly convex function. Hence, a good approximation of $\hat{B}_x$ is obtained by evaluating (18) on a reduced subset of voxels $\mathcal{V}_r \subset \mathcal{V}$, where $r$ is a user-defined subsampling factor. Then, using spline interpolation, we obtain an approximation of (18) whose evaluation is fast.

### E. Early echoes

As stated in Section II, the SWFT is only valid asymptotically, *i.e.* for modelling the late part $h^l$ of the RIR. This limitation is taken into consideration in the Taylor-SWFT algorithm, with early reflections $h^e$ being generated using a low-order ISM. In the end, the resulting output RIR is:

$$\hat{h}_x[t] = (1 - \varphi[t])h_x^e[t] + \lambda \varphi[t] h_x^l[t], \quad (19)$$

where $\lambda$ is a scaling factor introduced for normalization purposes and $\varphi$ is a cross-faded cosine profile

$$\varphi[t] = \frac{1}{2}\left(1 - \cos\left(\pi \frac{t - t_0}{t_m - t_0}\right)\right), \quad (20)$$

with $t_m$ and $t_0$ being the delays associated to, respectively, the last and first echoes. The scaling factor $\lambda$ is tuned in a similar way as in [7], so that the energy in $h_x^l$ and $h_x^e$ are matched around each echo. Formally, writing $h_x^e[t] = \sum_{j=0}^{m} e_j \delta_{t_j}[t]$ and choosing a small frame width $\tau > 0$, we define

$$E_j := e_j^2 \quad \text{and} \quad \hat{E}_j := \sum_{k=\lfloor F_s(t_j - \frac{\tau}{2}) \rfloor}^{\lfloor F_s(t_j + \frac{\tau}{2}) \rfloor} \left| h_x^l\left[\frac{k}{F_s}\right] \right|^2. \quad (21)$$

Then, $\lambda := \sqrt{\operatorname{median}_{j=0,\ldots,m}\{E_j/\hat{E}_j\}}$.

## IV. Experiments and discussion

The proposed algorithm is tested on the Benchmark for Room Acoustical Simulation (BRAS) [12] for its variety of complex acoustic scenes. The code used to obtain these results is available online[1].

### A. Experimental settings

The subset of BRAS [12] used in the experiments contains 209 RIRs measured in four different rooms: **coupled rooms** linked by an open door, a **seminar room** (small), a **music chamber hall** (medium) and an **auditorium** (large). Each room comes with a 3D model, material properties linked to the walls, source and microphone position and the measured RIRs. Other parameters like temperature and humidity are provided but not used here. Moreover, all experiments consider the microphone and the source to be omnidirectional. Finally, our experiments test the proposed algorithm Taylor-SWFT (T-SWFT) against the following baselines:

**ISM:** For each RIR in the dataset, the virtual images of the source are computed using the geometry of the room, up to a certain order of reflection $o$ (0 for direct path, 1 for one reflection on every wall, etc). This method is accurate, yet its computational cost grows exponentially with the order of reflection. Hence, ISM is mostly adapted for early reflections simulation and struggles to produce late reverberation in a reasonable time.

**RT:** For each RIR in the dataset, $n_{rt}$ rays are shot randomly in the room from the source. If their path reaches the microphone within a radius $r_{rt}$, then its contribution is added to the resulting RIR.

**ISM-RT:** This baseline is the default method implemented in [17] for RIRs generation. Early echoes are accurately captured using a low-order ISM, and late reverberation is synthesised via RT for efficiency. The resulting RIR is obtained as the sum of its late and early components.

**Gaussian noise:** Gaussian noise is sampled and multiplied by an exponential decay corresponding to the mean $RT_{60}$ of the room. The length of the generated RIR is chosen to reach 120 dB of attenuation.

### B. Choice of hyper-parameters

We chose 5000 rays for RT in order to have a long enough late reverberation and most of the early echoes. More rays led to similar results with longer computation times. The order of reflection of ISM is computed dynamically, given an upper number of *wanted sources* $W$ and the number of faces $F$ in the room, to get an upper limit of the computational time: $o = \lfloor \frac{\ln W}{\ln F} \rfloor$. For RIRs generated with ISM, to reach the late reverberation part, we chose $W = 1e10$. For T-SWFT, we use $W = 1e6$ because we only need the early echoes from ISM. The voxel size is set as 5% of the smallest dimension of the room, and the subsampling factor is chosen such that a point is computed every 8 voxels on average, to get enough precision without making the computational time excessive. The same parameters ($W = 1e6$, 5000 rays) are used in ISM-RT for consistency in the comparison with T-SWFT.

### C. Metrics

The RIRs generated using the different methods is compared with the measured ones using several standard acoustic metrics:
- Clarity 50 ms ($C_{50}$),
- Definition 50 ms ($D_{50}$),
- Reverberation Time 30 dB ($RT_{30}$),
- Energy Decay Relief (EDR),
- Energy Decay Curve (EDC).

These metrics are used to compute objective attributes that are compared to the original ones with the following distance function:

$$\Delta_{\text{metric}} = \|\text{metric}(\text{ref}) - \text{metric}(\text{generated})\|_1. \quad (22)$$

---
[1] https://github.com/TAYLOR-SWFT/Taylor-SWFT

TABLE I
Comparison of the methods. Each cell corresponds to the mean ± standard deviation over 209 RIRs. The lower is better for all metrics.

|  | Coupled Rooms | Seminar Room | Chamber Music Hall | Auditorium |
|---|---|---|---|---|
| ↓ $\Delta D_{50}$ (%) | | | | |
| T-SWFT | $20.1 \pm 18.5$ | $7.21 \pm 4.73$ | $19.2 \pm 17.4$ | $15.3 \pm 16.1$ |
| ISM-RT | $14.4 \pm 11.9$ | $9.86 \pm 13$ | $24.8 \pm 20.1$ | $39.3 \pm 29.4$ |
| RT | $19 \pm 12.5$ | $15.5 \pm 6.7$ | $30.9 \pm 23.7$ | $38.4 \pm 28.3$ |
| ISM | $53.4 \pm 24.7$ | $57.8 \pm 6.9$ | $32.1 \pm 18$ | $16 \pm 12.1$ |
| Noise | $21 \pm 15.6$ | $8.1 \pm 6.3$ | $22.3 \pm 13.3$ | $26.4 \pm 12.9$ |
| ↓ $\Delta RT_{30}$ (s) | | | | |
| T-SWFT | $0.94 \pm 0.77$ | $0.06 \pm 0.03$ | $0.23 \pm 0.04$ | $0.07 \pm 0.06$ |
| ISM-RT | $0.73 \pm 0.5$ | $0.15 \pm 0.04$ | $0.39 \pm 0.13$ | $0.81 \pm 0.24$ |
| RT | $0.8 \pm 0.49$ | $0.16 \pm 0.04$ | $0.38 \pm 0.12$ | $0.78 \pm 0.27$ |
| ISM | $1.66 \pm 1.1$ | $0.82 \pm 0.03$ | $0.53 \pm 0.04$ | $0.96 \pm 0.09$ |
| Noise | $1.35 \pm 1.08$ | $0.23 \pm 0.03$ | $0.17 \pm 0.03$ | $0.64 \pm 0.08$ |
| ↓ $\Delta EDC$ (dB) | | | | |
| T-SWFT | $28.1 \pm 8$ | $18.3 \pm 2.1$ | $22.2 \pm 1.4$ | $20.2 \pm 1.7$ |
| ISM-RT | $12.9 \pm 3.6$ | $7.38 \pm 2.21$ | $16.2 \pm 5.3$ | $30.4 \pm 6.4$ |
| RT | $13.3 \pm 4.1$ | $6.85 \pm 0.83$ | $16.1 \pm 4.8$ | $29.3 \pm 7.5$ |
| ISM | $62.3 \pm 9.3$ | $56.3 \pm 2.1$ | $52 \pm 3.2$ | $59.2 \pm 2.6$ |
| Noise | $50.7 \pm 8.9$ | $26.9 \pm 2$ | $33.1 \pm 3.2$ | $42.7 \pm 2.7$ |
| ↓ $\Delta EDR$ (dB) | | | | |
| T-SWFT | $17.5 \pm 4.9$ | $12.1 \pm 1.5$ | $13.1 \pm 2.4$ | $14.9 \pm 2.1$ |
| ISM-RT | $11.9 \pm 1.9$ | $9.99 \pm 0.89$ | $13.7 \pm 2.1$ | $23 \pm 4.1$ |
| RT | $11.9 \pm 2.2$ | $9.86 \pm 0.58$ | $13.6 \pm 1.6$ | $22.5 \pm 4.9$ |
| ISM | $27.3 \pm 5.5$ | $25.1 \pm 2$ | $22.7 \pm 3.5$ | $28.6 \pm 2.6$ |
| Noise | $22.4 \pm 4.9$ | $14.9 \pm 1.5$ | $16.3 \pm 2.9$ | $22.4 \pm 2.3$ |
| ↓ DTW | | | | |
| T-SWFT | $7.57 \pm 6.07$ | $5.12 \pm 1.89$ | $4.27 \pm 1.22$ | $4.08 \pm 1.6$ |
| ISM-RT | $8.16 \pm 5.76$ | $6.43 \pm 1.63$ | $5.23 \pm 1.73$ | $6.96 \pm 2.66$ |
| RT | $9.92 \pm 3.03$ | $13.6 \pm 3$ | $5.84 \pm 1.48$ | $6.86 \pm 2.64$ |
| ISM | $9.7 \pm 8.19$ | $6.55 \pm 2.35$ | $4.91 \pm 2.29$ | $4.99 \pm 2.69$ |
| Noise | $29.5 \pm 1.7$ | $42.1 \pm 1.9$ | $37 \pm 1.5$ | $36.2 \pm 1.4$ |
| ↓ Computation Time (s) | | | | |
| T-SWFT | $0.74 \pm 0.16$ | $0.85 \pm 0.03$ | $0.66 \pm 0.03$ | $0.92 \pm 0.03$ |
| ISM-RT | $13.9 \pm 1.5$ | $18.2 \pm 0.1$ | $51.9 \pm 0.3$ | $61.6 \pm 0.2$ |
| RT | $13.8 \pm 1.5$ | $18 \pm 0$ | $51.9 \pm 0.3$ | $61.5 \pm 0.2$ |
| ISM | $413 \pm 414$ | $52.5 \pm 3.6$ | $27.7 \pm 0.5$ | $0.27 \pm 0.01$ |
| Noise | $0 \pm 0$ | $0 \pm 0$ | $0 \pm 0$ | $0 \pm 0$ |

We also included Dynamic Time Warping (DTW) [18], which computes all the possible combinations of alignment between two time series and finds the cheapest path to align both signals. We chose this metric to account for the early echoes, which contain a lot of energy.

*D. Results and discussion*

Table I shows the results grouped by room and metrics, with the best and second best values of each group being highlighted. Each tested room represents a specific challenge for the evaluated models as described in the following. The coupled rooms are connected with a door, and RIRs were measured with the door opened at two different angles. Coupled rooms are difficult to model due to the resonance of the second room and the filtering effects of the connection. The seminar room has a complex $RT_{60}$ profile in the low frequencies, where the SWFT is not completely valid. The chamber music hall presents a closed, empty volume that stands on top of it, thereby constituting another case of coupled rooms. Lastly, the auditorium is a large room with seats, a stage, and diverse furniture. Among all rooms, the auditorium most closely conforms to the SWFT's assumptions.

As expected, our method T-SWFT performs best in the auditorium, with a good estimation of the acoustic parameters. However, it fails to model the coupled rooms and is slightly off in the chamber music hall. The SWFT equations use the volume of the two parts of the coupled rooms without taking into account the connection, which only lets through a fraction of the sound, resulting in wrong estimations. The seminar room is well handled by T-SWFT overall, but the EDC and EDR are worse than other methods, especially RT. This can be explained by the complex low-frequency profile of the room, which falls outside the scope of the SWFT. The most important part of the results is the computation time. We measured the computation time required for all methods to compute a RIR for one source and one microphone at given positions. The method with exponentially decaying noise runs in less than 1 ms per run, but it comes with the worst early echoes prediction. Then, T-SWFT is the fastest method among the others while producing accurate RIRs, with good temporal alignment as shown with DTW, precise $RT_{30}$ estimation in non-coupled rooms, and overall accurate estimation of EDR and EDC.

For comparison consistency, the computation times include the full generation of $\hat{h}_x$ (ISM, cross-fade and late reverberation). However, in a real-time scenario, part of the late reverberation can be computed during pre-processing, resulting in a much more efficient algorithm (see Section III.B). We measured the real-time ratio $t_p/t_r$ with $t_p$ being the time required to apply reverberation on $t_r$ seconds of audio of our method. 490 chunks of 3-second audio were reverberated in the seminar room using RIRs generated by T-SWFT, with an ISM order of 1, a sampling rate of 16 kHz and a buffer size of 40 ms. On average, the real-time ratio was of 0.698 with a standard deviation of 0.033 using an Intel Core Ultra 9 285. The ratio being consistently under 1 means T-SWFT can be run in real-time with appropriate implementation and hardware.

V. CONCLUSION

In this paper, we presented an efficient method for late reverberation synthesis based on the Statistical Wave Field Theory (SWFT). The proposed approach, Taylor-SWFT (T-SWFT), enables real-time computation of late reverberation for dynamic configurations involving moving sources and microphones, which, to the best of our knowledge, has not been previously achieved. Furthermore, it can be combined with fast early reflection models, such as low-order ISM. By enabling real-time reverberation in dynamic scenes, T-SWFT has the potential to enhance immersion in spatial audio applications, including virtual reality and video games. Finally, future work will focus on extending the method to coupled rooms, improving its accuracy in the low-frequency regime, and incorporating source-position dependence in (18). Another interesting research direction lies in the analysis of the family of parameters defined by (12) to establish formal connections to the SWFT.

## References

[1] L. Savioja and U. P. Svensson, "Overview of geometrical room acoustic modeling techniques," *J. Acoust. Soc. Am.*, vol. 138, no. 2, pp. 708–730, 2015, doi: 10.1121/1.4926438.

[2] C. Schissler and D. Manocha, "GSound: Interactive Sound Propagation for Games," *Proceedings of the AES International Conference*, p. , 2011.

[3] R. L. Pedersen, L. Picinali, N. Kajs, and F. Patou, "Virtual-Reality-Based Research in Hearing Science: A Platforming Approach," *Journal of the Audio Engineering Society*, vol. 71, no. 6, pp. 374–389, June 2023, doi: https://doi.org/10.17743/jaes.2022.0083.

[4] S. Giacomelli *et al.*, "Remote Immersive Audio Production: State of the Art Implementation, Challenges, and Improvements," in *2024 IEEE 5th International Symposium on the Internet of Sounds (IS2)*, 2024, pp. 1–10. doi: 10.1109/IS262782.2024.10704192.

[5] A. Krokstad, S. Strom, and S. Sørsdal, "Calculating the acoustical room response by the use of a ray tracing technique," *Journal of Sound and Vibration*, vol. 8, no. 1, pp. 118–125, 1968.

[6] S. Siltanen, T. Lokki, S. Kiminki, and L. Savioja, "The room acoustic rendering equation," *J. Acoust. Soc. Am.*, vol. 122, no. 3, pp. 1624–1635, 2007.

[7] E. A. Lehmann and A. M. Johansson, "Diffuse Reverberation Model for Efficient Image-Source Simulation of Room Impulse Responses," *IEEE Trans. Audio, Speech, Lang. Process.*, vol. 18, no. 6, pp. 1429–1439, 2010, doi: 10.1109/TASL.2009.2035038.

[8] Y. Masuyama, F. Germain, G. Wichern, C. Ick, and J. Le Roux, "Physics-Informed Direction-Aware Neural Acoustic Fields," in *Proc. WASPAA*, 2025, pp. 1–5. doi: 10.1109/WASPAA66052.2025.11230918.

[9] X. Liu *et al.*, "Hearing Anywhere in Any Environment," in *Proceedings of the IEEE/CVF Conference on Computer Vision and Pattern Recognition (CVPR)*, June 2025, pp. 5732–5741.

[10] L. Kelley, D. Di Carlo, A. A. Nugraha, M. Fontaine, Y. Bando, and K. Yoshii, "RIR-in-a-Box: Estimating room acoustics from 3D mesh data through shoebox approximation," in *Interspeech*, 2024.

[11] R. Badeau, "Statistical wave field theory," *J. Acoust. Soc. Am.*, vol. 156, no. 1, pp. 573–599, July 2024, doi: 10.1121/10.0027914.

[12] F. Brinkmann, L. Aspöck, D. Ackermann, R. Opdam, M. Vorländer, and S. Weinzierl, "A Benchmark for Room Acoustical Simulation. Concept and Database," *Applied Acoustics*, vol. 176, p. 107867, May 2021, doi: 10.1016/j.apacoust.2020.107867.

[13] J.-D. Polack, "La transmission de l'energie sonore dans les salles (Transmission of sound energy in rooms)," Doctoral dissertation, 1988. [Online]. Available: http://www.theses.fr/1988LEMA1011

[14] A. G. Prinn and R. Badeau, "Verification of reverberation time predictions derived from the statistical wave field theory," *Applied Acoustics*, vol. 250, p. 111337, June 2026.

[15] A. Aknin and R. Badeau, "Stochastic reverberation model with a frequency dependent attenuation," in *Proc. WASPAA*, 2021, pp. 351–355.

[16] A. Aknin and R. Badeau, "Algorithmes rapides pour la modélisation d'une réponse de salle dont l'atténuation dépend de la fréquence (Fast algorithms for modeling a room response whose attenuation depends on frequency)," in *16e Congrès Français d'Acoustique (CFA 2022)*, Marseille, France, Apr. 2022. [Online]. Available: https://telecom-paris.hal.science/hal-03559398

[17] R. Scheibler, E. Bezzam, and I. Dokmanić, "Pyroomacoustics: A Python Package for Audio Room Simulation and Array Processing Algorithms," in *2018 IEEE International Conference on Acoustics, Speech and Signal Processing (ICASSP)*, 2018, pp. 351–355. doi: 10.1109/ICASSP.2018.8461310.

[18] M. Cuturi and M. Blondel, "Soft-DTW: A Differentiable Loss Function for Time-Series," in *Proceedings of the 34th International Conference on Machine Learning*, PMLR, July 2017, pp. 894–903.